\documentclass[onecolumn,amsmath,amssymb,aps,showpacs,superscriptaddress]{revtex4}

\usepackage[dvips]{graphicx}
\usepackage[dvips]{color}
\usepackage{hyperref,breakurl,amsmath,amssymb,pst-node,eepic}
\usepackage{graphicx}
\usepackage{dcolumn}
\usepackage{bm}
\usepackage{hyperref}

\begin{document}

\title{What do leaders know?}

\author{Giacomo Livan}
\email{glivan@ictp.it}
\affiliation{Abdus Salam International Centre for Theoretical Physics, Strada Costiera 11, 34151 Trieste, Italy}
\author{Matteo Marsili}
\email{marsili@ictp.it}
\affiliation{Abdus Salam International Centre for Theoretical Physics, Strada Costiera 11, 34151 Trieste, Italy}

\date{\today}

\begin{abstract}
The ability of a society to make the right decisions on relevant matters relies on its capability to properly aggregate the noisy information spread across the individuals it is made of. In this paper we study the information aggregation performance of a stylized model of a society whose most influential individuals -- the leaders -- are highly connected among themselves and uninformed. Agents update their state of knowledge in a Bayesian manner by listening to their neighbors. We find analytical and numerical evidence of a transition, as a function of the noise level in the information initially available to agents, from a regime where information is correctly aggregated to one where the population reaches consensus on the wrong outcome with finite probability. Furthermore, information aggregation depends in a non-trivial manner on the relative size of the clique of leaders, with the limit of a vanishingly small clique being singular.
\end{abstract}

\maketitle

\begin{quotation}
The Chinese famines of 1958-1961 killed, it is now estimated, close to thirty million people [...]. 
The so called Great Leap Forward initiated in the late 1950s had been a massive failure, but the Chinese government refused to admit that and continued to pursue dogmatically much of the same disastrous policies for three more years [...]. In 1962, just after the famine had killed so many millions, Mao made the following observation, to a gathering of seven thousand cadres:
{\em ``Without democracy, you have no understanding of what is happening down below; the situation will be unclear; you will be unable to collect sufficient opinions from all sides; there can be no communication between top and bottom; top-level organs of leadership will depend on one sided and incorrect material to decide issues [...]''.} \\
\rightline{(from A. Sen \cite{Sen})~~~~~~~~~{}}
\end{quotation}

\section{Introduction}
\label{intro}

Amartya Sen \cite{Sen} argues that famine and other catastrophes are easily avoided in a democracy. This argument relies on the fact that where information can freely diffuse, decision makers can form an unbiased picture of the state of a society, and take proper measures. Biases due to individual opinions are expected to be washed out in the information aggregation process, a phenomenon often referred to as the ``wisdom of crowds'' \cite{Surowiecki}. 
Still, cases of information aggregation failure abound even in democratic societies\footnote{While writing, Turkey is in  a state of turmoil exacerbated by the fact that the democratically elected government failed to properly understand and respond to the issues that were raised \cite{Shafak}.}. For example, in the aftermath of the 2008 Lehman Brothers bankruptcy, former Federal Reserve chairman Alan Greenspan expressed his state of ``shocked disbelief'' during his hearing before the US Committee of Government Oversight and Reform, leaving the public opinion to wonder where did he get the information the policy of the FED was based on. Al Gore \cite{Gore} argues that shortcuts between decision makers and the media are often such that the former are not in the best position to be informed about what is going on.

A number of models on social dynamics have addressed the issue of information aggregation (see e.g. \cite{Castellano} for an excellent review). The simplest is probably the voter model, which entails agents taking the same opinion of a randomly chosen node amongst their neighbors. This allows for sharp predictions \cite{Clifford,Redner} where generally the information aggregation process converges to the incorrect outcome with finite probability. Other contributions have instead proposed different opinion dynamics mechanisms, such as majority rules \cite{Galam} or social impact models \cite{Sznajd}, which support different conclusions. These models, however, come short in their micro-economic foundation, as the interaction mechanism is somewhat arbitrary.

More detailed micro-economic models of social learning have been proposed in the Economics literature. It is well known that information aggregation may fail when agents free ride on the information gathered by others, without seeking independent sources. This phenomenon, called rational herding \cite{herding}, is also supported by experimental evidence \cite{Lorenz}.
 
A sequel of papers have focused on Bayesian learning schemes\footnote{At odds with models of rational herding, where agents deduce signals from the behavior of others, in Bayesian learning schemes agents exchange the full probability distribution on the signals they have received. This avoids the loss of information which lies at the heart of rational herding. This phenomenon in the context of a social network is discussed e.g. in \cite{curty}.} \cite{DeGroot,Gale,Duffie1,Duffie2}, coming to the generic conclusion that when agents update their beliefs following Bayes rule society correctly aggregates information (still see \cite{Acemoglu,Marsili,Smith}). Some authors have focused on the impact of dominant groups of individuals  on the aggregation of information. For example, Bala and Goyal \cite{Bala} introduced the notion of ``Royal Family'' as a group of agents whose behavior is observed by anyone else. Alternatively, Golub and Jackson \cite{Golub} defined $t$-step ``prominent groups'' as those groups whose behavior eventually influences all other agents within time $t$. Regardless of the specific definitions, these and other studies unanimously highlighted the negative role that exceedingly influential groups have on the information aggregation process. 

In this paper we focus on an extremely stylized model of a society and we address the issue of whether information distributed across the population is able to diffuse to an uniformed well connected clique of decision makers. Our model assumes Bayesian learning, but differently from \cite{Duffie1,Duffie2}, who study a continuum of agents, we study a finite but large population of agents connected by a social network. On a finite network, when agents talk repeatedly with their peers, they may not be able to disentangle what in their peer's opinion is new information and what reflects information exchanged in previous interactions, including the one provided by themselves to them. This phenomenon, called ``persuasion bias'' in \cite{Demarzo}, introduces a non-trivial positive feedback and leads to information aggregation failure, at odds with the conclusions of \cite{Duffie1,Duffie2}. 

The main conclusions of our paper can be summarized in two points: {\em i)} information aggregation crucially depends on the synchronicity of the information updates of different agents: in the extreme case of a parallel update dynamics, where we can derive analytic results, information diffusion leads to the correct outcome in the limit of a very large society or for very informative initial signals. When the fraction of agents who update their beliefs at each time step is lower than a critical threshold, the society converges with finite probability to the wrong outcome, no matter how large the society is. {\em ii)} In the case of parallel dynamics, information aggregation {\em degrades} as the size of the clique of uninformed agents gets smaller. In particular, the limit of a vanishingly small clique of uninformed agents, behaves markedly differently from the case of a homogeneous society (with no clique). Both results suggests that it might not be wise to rely on crowds in situations which are reminiscent of those prevailing in our societies, 
where update is sequential and the social network is characterized by highly connected cliques (news corporations, political parties). 

The paper is organized as follows. In Section \ref{dyn_topology} we detail the network structure and the information update rules, as well as our quantitative measure of a social network's ability to correctly aggregate information. In Section \ref{prob_computing} we provide analytic results for the case of parallel information update. In Section \ref{diff_dyn} we numerically compare these results with those obtained in sequential update schemes. We conclude the paper with a few conclusive remarks in Section \ref{conclusions}.

\section{Dynamics and network topology}
\label{dyn_topology}

Following Ref. \cite{Duffie1,Duffie2}, let us consider a population of agents who need to decide whether a certain event $X$ will occur ($X = +1$) or not ($X = -1$). Let us denote the prior probability of $X$ as $P_0 \{ X = +1 \} = \nu = 1 - P_0 \{ X = -1 \}$, where $\nu \in (0,1)$. 

\subsection{Core-periphery network structure}

As already stated, our interest is mainly focused on the information aggregation process as performed by societies where a fraction of  individuals matters much more than the vast majority of the population. In the language of networks, the most obvious measure of the importance of a node is its degree, i.e. the number of neighbors. For this very reason, throughout the rest of the paper we shall focus on a highly stylized society structure, where only few nodes have a large degree, which we build starting from a connected regular graph where all $N$ nodes have degree $c \sim \mathcal{O}(1)$. Then, a randomly chosen set $\mathcal{H}$ made of $N_\mathcal{H}$ non-neighboring sites are connected among themselves, thus forming a clique of nodes with degree $N_\mathcal{H} + c -1$ (this construction is such that each hub has exactly $c$ links connecting it to nodes outside $\mathcal{H}$). In the following, we shall be mostly interested in the case $N_\mathcal{H} \gg c$, i.e. when $\mathcal{H}$ becomes a group of mutually connected hubs.

\subsection{Initial beliefs}

At time $t = 0$, each agent $i \notin \mathcal{H}$ receives signals about event $X$ which are independently drawn from a probability distribution  $P_{\not\mathcal{H}} \{ \mathbf{s} | X \}$. We assume these signals to be informative \cite{Duffie1,Acemoglu}, i.e.
$P_{\not\mathcal{H}}  \{ s_i = \pm 1 | X = \pm 1 \} > P_{\not\mathcal{H}}  \{ s_i = \pm 1 | X = \mp 1  \}$ $\forall i$, and we focus on the particular case 
\begin{equation} \label{info_sig}
p = P_{\not\mathcal{H}}  \{ s = X | X \}=1-P_{\not\mathcal{H}}  \{ s = -X | X \}.
\end{equation}
On the other hand, the agents $i\in\mathcal{H}$ -- the leaders -- are assumed to be initially uninformed. This means that their signals are independently drawn from a probability $P_{\mathcal{H}}  \{ s_i | X\}=1/2$ for $s_i,X=\pm 1$.

\subsection{Belief update dynamics}

In our model, agents repeatedly exchange information with their neighbors. In this exchange, the generic agent $i$ collects a certain number $n$ of signals that we denote by $\mathbf{s}_i= (s^{(0)}_i,s^{(1)}_i \ldots, s^{(n)}_i)$, where $s^{(0)}_i$ are the initial signals discussed above. Given this information set $\mathbf{s}_i$, by Bayes' theorem \cite{Lee}, the agent's state of knowledge about $X$ is quantified by the conditional probability
\begin{equation} \label{Bayes}
P \{ X | \mathbf{s}_i \} = \frac{P \{ \mathbf{s}_i | X \} P_0 \{ X \} }{P \{ \mathbf{s}_i \}}  \ ,
\end{equation}
where $P \{ \mathbf{s}_i \}$ is the probability of the signals $\mathbf{s}_i$. Notice that the likelihood ratio of $P \{ X = +1 | \mathbf{s}_i \}$ and $P \{ X = -1 | \mathbf{s}_i \}$ does not depend on $P \{ \mathbf{s}_i \}$. 
If the agent believes that the different signals are independent, then 
\begin{equation}
P \{ \mathbf{s}_i | X \} = \prod_{a=0}^n P \{ s_i^{(a)} | X \} \ ,
\end{equation}
and the logarithm of the likelihood ratio, which embodies the state of information of agent $i$, can be described by a single variable $\theta_i$:
\begin{equation} \label{theta}
\theta_i = \log \frac{P \{ X = +1 | \mathbf{s}_i \}}{P \{ X = -1 | \mathbf{s}_i \}} - \log \frac{\nu}{1 - \nu} = \sum_{a=0}^n \log \frac{ P \{ s_i^{(a)} | X = +1 \}}{P \{ s_i^{(a)} | X = -1 \}} \ .
\end{equation}

At $t=0$, agents have just one signal. Then we have $n = 0$ and the above expression reduces to the very compact form
\begin{equation} \label{compact_theta}
\theta_i = s_i^{(0)} \log \frac{p}{1-p}.
\end{equation}

When two agents, say $i$ and $j$ with signals $\mathbf{s}_i$ and $\mathbf{s}_j$ respectively, meet, they communicate by exchanging signals and, as a result, their state of knowledge changes. Indeed, if $\mathbf{s}_i \rightarrow \mathbf{s}_i^\prime = (\mathbf{s}_i, \mathbf{s}_j)$, then $\theta_i \rightarrow \theta_i^\prime = \theta_i + \theta_j$. Likewise, if $\mathbf{s}_j \rightarrow \mathbf{s}_j^\prime = (\mathbf{s}_i, \mathbf{s}_j)$, then $\theta_j \rightarrow \theta_j^\prime = \theta_i + \theta_j$.

Starting from an initial state of knowledge $\theta_i (t = 0)$, for $i = 1, \ldots, N$, one can think of different types of information update. Our assumption will be that at each time step $t = 1, 2, \ldots$, a certain fraction $\Phi = N_\Phi/ N$ (where $N_\Phi \leq N$) of randomly selected agents update their state of knowledge by \emph{listening} to their neighbors. So, assuming that agents in the set $\mathcal{I}_t = \{ i_1, i_2, \ldots, i_{N_\Phi} \} $ are the ones to update their information at time $t$, one has:
\begin{equation} \label{general_dyn}
\theta_i (t+1) = \theta_i(t) + \sum_{j = 1}^N a_{ij} \theta_j (t) \ , \ \forall \ i \in \mathcal{I}_t \ ,
\end{equation}
where $a_{ij}$ is the $(i,j)$ element of the adjacency matrix $A = \{ a_{ij} \}_{i,j = 1, \ldots, N}$, i.e. 
$a_{ij} = a_{ji} = 1$ if agents $i$ and $j$ are connected and $a_{ij} = a_{ji} = 0$ if they are not. 

Clearly, the above dynamics has two limiting cases: $\Phi = 1 / N$ and $\Phi = 1$. The former describes cases where agents update their information one at a time, and we shall refer to this particular situation as random node sequential (RNS) dynamics. The latter case, instead, describes a parallel dynamics where all agents simultaneously update their state of knowledge. This information update rule was initially proposed in \cite{DeGroot}, and, due to its analytical tractability, represents the most frequent choice in social learning models. In the following, we also shall investigate this type of dynamics, and then explore other cases in Section \ref{diff_dyn}.

The dynamics in Eq. (\ref{general_dyn}) is unbounded, i.e. each $\theta_i$ will either diverge to $+ \infty$ or $- \infty$. Thus, information aggregation properties can be assessed simply by looking at the signs of the $\theta_i$s in the long run. Thus, a good measure of information aggregation is given by the ``magnetization'' of the system:
\begin{equation} \label{magnetization}
\Theta(t) = \frac{1}{N} \sum_{i=1}^N \mathrm{sign} (\theta_i(t)) \ .
\end{equation}
The quantity $X \Theta(t)$ tells what is the fraction of the population holding the right information on event $X$ at time $t$. A quantitative measure of information aggregation is given by the probability $P \{ X \Theta(t) > 0 \}$ that the majority will converge to the true outcome, in an ensemble of repeated trials.

\section{Parallel dynamics}
\label{prob_computing}

According to the parallel dynamics prescription, all agents in a social network listen to their neighbors at any time $t = 1, 2 \ldots$, and update their state of knowledge accordingly:
\begin{equation} \label{dynamics}
\theta_i (t+1) = \theta_i(t) + \sum_{j = 1}^N a_{ij} \theta_j (t) \ , \ \forall \ i \ .
\end{equation}
By collecting all $\theta_i(t)$s into a column vector $| \theta(t) \rangle$\footnote{Here we switch to bra / ket notation, i.e. we denote by $| w \rangle$ the column vector with components $w_1, w_2, \ldots$, and by $\langle w |$ the corresponding row vector.}, the dynamics described in equation \eqref{dynamics} can be rewritten as 
\begin{equation} \label{dyn_evolution}
| \theta(t) \rangle = (1+ A)^t | \theta (0) \rangle \ .
\end{equation}
The above equation clearly suggests that the spectral properties of the adjacency matrix $A$ play a crucial role in the time evolution of the state of knowledge vector $| \theta(t) \rangle$. Being symmetric, the adjacency matrix $A$ yields $N$ real eigenvalues $\lambda_1 \geq \lambda_2 \geq \ldots \geq \lambda_N$, whose corresponding eigenvectors $| \lambda_i \rangle$ ($i = 1, \ldots, N$) form an orthogonal set in $\mathbb{R}^N$. By decomposing the adjacency matrix as $A = \sum_{i=1}^N \lambda_i | \lambda_i \rangle \langle \lambda_i |$, one can see that, for large enough times, equation \eqref{dyn_evolution} becomes
\begin{equation} \label{dyn_evolution_2}
| \theta(t) \rangle \simeq (1 + \lambda_1)^t \langle \lambda_1 | \theta(0) \rangle | \lambda_1 \rangle \ .
\end{equation}
As is well known from Frobenius-Perron theorem \cite{Berman}, all components of the eigenvector $| \lambda_1 \rangle$, corresponding to the largest eigenvalue of the adjacency matrix $A$, share the same sign, which we shall assume to be positive from now on. Thus, in the light of the relation in \eqref{dyn_evolution_2}, two main points become apparent:
\begin{itemize}
	\item For large enough times $| \theta(t) \rangle$ is proportional to $| \lambda_1 \rangle$, meaning that all agents on the network either learn the correct value of $X$ or they all get it wrong.
	\item The sign of the components in $| \lambda_1 \rangle$ is completely determined by the sign of the overlap $\langle \lambda_1 | \theta (0) \rangle$, so that the probability of the whole network learning the right information reads
	\begin{equation} \label{sc_prod}
	P \{ X \Theta(t) > 0 \} = P \{ X \langle \lambda_1 | \theta (0) \rangle > 0 \} \ .
	\end{equation}
\end{itemize} 
In the following we shall compute the probability \eqref{sc_prod} for the simple network topology discussed above.

\begin{figure}
\begin{center}
\resizebox{0.6\columnwidth}{!}{
  \includegraphics{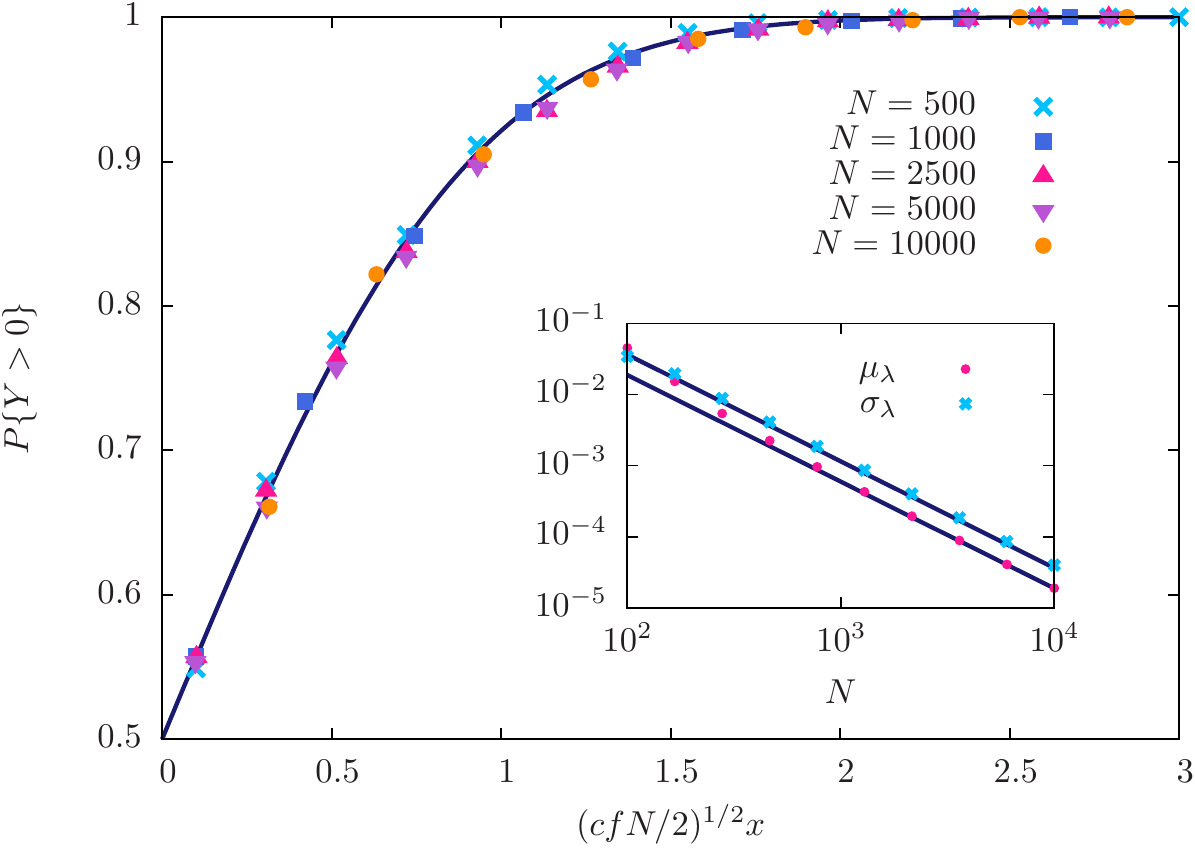}}
\end{center}
\caption{The prediction by equation \eqref{prob_Y_3} for the probability of correct information aggregation (solid line) is compared with the results of numerical simulations run with the parallel dynamics of equation \eqref{dyn_evolution}. Each dot represents the empirical estimate of such a probability for a given value of $x$ computed as the fraction (over $10^4$ samples) of networks that reached consensus on the true value of $X$. All simulations were performed on networks with $c=4$ and $f=0.05$ for different values of the system size $N$ (shown in the plot). INSET: Comparison between the large $N$ approximations (solid lines) for $\mu_\lambda$ and $\sigma_\lambda$ in equation (19) and the corresponding quantities estimated by averaging over the top eigenvectors $| \lambda_1 \rangle$ of 100 network configurations. As can be seen, for large enough values of $N$ the empirically measured mean and standard deviation are in excellent agreement with the approximations in equation \eqref{stat_lambda}. In all cases we have $c = 4$ and $f = 0.05$.}
\label{ls}
\end{figure}

For the sake of simplicity, let us assume $X = +1$, so that the probability in equation \eqref{sc_prod} is equivalent to the probability of the scalar product $\langle \lambda_1 | \theta(0) \rangle$ being positive, and that each agent is initially given one signal $s = \pm 1$ at time $t = 0$. Assuming that hubs, i.e. nodes in the clique $\mathcal{H}$, have no initial information ($\theta_i (0) = 0$ for $i \in \mathcal{H}$), such a scalar product can be written as a sum over the $N - N_\mathcal{H}$ sites not belonging to $\mathcal{H}$:
\begin{equation} \label{sc_prod_2}
Y = \langle \lambda_1 | \theta(0) \rangle = \sum_{i \notin \mathcal{H}} \lambda_1^{(i)} \theta_i (0) \ ,
\end{equation}
where $\lambda_1^{(i)}$ denotes the $i$-th component of the first eigenvector, and $\theta_i (0) = s_i \log \frac{p}{1-p}$ (see equation \eqref{compact_theta}). A good approximation scheme to estimate the probability of the quantity in equation \eqref{sc_prod_2} being positive is via the central limit theorem: as a matter of fact the scalar product in \eqref{sc_prod_2} is the sum of $N - N_\mathcal{H}$ random variables, each given by the product of two random variables: $y_i = \lambda_1^{(i)} \theta_i(0)$. Thus, the probability of $Y$ in equation \eqref{sc_prod_2} being positive is approximately given by
\begin{equation} \label{prob_Y}
P \{ Y > 0 \} \simeq \frac{1}{2} \text{erfc} \left ( - \frac{\mu_Y}{\sqrt{2} \sigma_Y} \right ) \ ,
\end{equation}
where $\mu_Y$ and $\sigma_Y$ denote the mean and standard deviation, respectively, of the random variable $Y$. Given the independence of the $\theta_i$s and the eigenvector components $\lambda_1^{(i)}$s, such two quantities are given by
\begin{eqnarray} \label{stat_Y}
\mu_Y &=& (N - N_\mathcal{H}) \mu_\theta \mu_\lambda \\ \nonumber
\sigma_Y^2 &=& (N - N_\mathcal{H}) \left ( \mu_\theta^2 \sigma_\lambda^2 + \mu_\lambda^2 \sigma_\theta^2 + \sigma_\theta^2 \sigma_\lambda^2 \right ) \ ,
\end{eqnarray}
where $\mu_\theta$ and $\sigma_\theta$ denote the mean and standard deviation of the random variables $\theta_i$, whereas $\mu_\lambda$ and $\sigma_\lambda$ denote the mean and standard deviation of the eigenvector components $\lambda_1^{(i)}$ for $i \notin \mathcal{H}$. Computing $\mu_\theta$ and $\sigma_\theta$ is easy. Recalling that signals must be informative (see equation \eqref{info_sig}), one has $p = P \{ s = +1 | X = +1 \} > 1/2$. Let us rewrite such probability as $p = (1+x)/2$ with $x \in (0,1)$. Then, one can immediately verify that
\begin{eqnarray} \label{stat_theta}
\mu_\theta &=& x \log \frac{1+x}{1-x} \\ \nonumber
\sigma_\theta &=& \sqrt{1-x^2} \log \frac{1+x}{1-x} \ .
\end{eqnarray}
As regards $\mu_\lambda$ and $\sigma_\lambda$, good approximate expressions for them can be computed by employing standard perturbation theory up to second order (see Appendix \ref{perturb} for the details). To leading order in $N$ one gets:
\begin{eqnarray} \label{stat_lambda}
\mu_\lambda &\simeq& \frac{c}{\sqrt{f} (1-f) N^{3/2}} \\ \nonumber
\sigma_\lambda^2 &\simeq& \frac{c (1 - f(c+1))}{f^2(1-f)^2 N^3} \ ,
\end{eqnarray}
where $f = N_\mathcal{H} / N$ denotes the fraction of hubs in the network. As can be seen from the inset in Fig. \ref{ls}, the above approximations are in excellent agreement with results obtained from numerical diagonalization of adjacency matrices, especially for large network sizes.

Plugging equations \eqref{stat_theta} and \eqref{stat_lambda} into equation \eqref{stat_Y}, one can eventually compute the probability of converging to the right value of event $X$ as in equation \eqref{prob_Y}:
\begin{equation} \label{prob_Y_2}
P \{ Y > 0 \} \simeq \frac{1}{2} \text{erfc} \left ( - x \sqrt{\frac{N}{2} \frac{cf(1-f)}{1-f(1+cx^2)}} \right ) \ .
\end{equation}
As already stated, we are mostly interested in cases where only a few nodes in the network play the role of hubs, i.e. $f \ll 1$: in this case the probability in equation \eqref{prob_Y_2} further simplifies to the following remarkably simple expression:
\begin{equation} \label{prob_Y_3}
P \{ Y > 0 \} \simeq \frac{1}{2} \text{erfc} \left ( - x \sqrt{\frac{N}{2} f c} \right ) \ .
\end{equation}
In Fig. \ref{ls} the prediction by the above equation is compared with the results of numerical simulations for $c = 4$, $f = 0.05$, and for several different system sizes $N$: all results are in very good agreement with equation \eqref{prob_Y_3} (all data points are rescaled in order to collapse on the function $\text{erfc}(-x) / 2$).

A few comments are in order on the approximate result of equation \eqref{prob_Y_3}. Since $\text{erfc}(-2)/2 \simeq 1$, according to equation \eqref{prob_Y_3} for each system size $N$ correct information aggregation happens with probability that for all practical purposes can be considered equal to 1 when initial signals' informativeness is $p \ge p_0 = (1+x_0)/2$, where $x_0 = 2 \sqrt{2/(Nfc)}$. This point essentially means that for \emph{any} population size $N$ correct information aggregation is possible, for informative enough initial signals, despite the presence of a fraction $f$ of dominant nodes. Such a result shows that the presence of a group of individuals with large influence does not necessarily jeopardize correct information aggregation. Moreover, the threshold value $x_0$ is inversely proportional to $\sqrt{N}$, meaning that large populations will be able to aggregate information correctly as soon as signals are informative, i.e. as soon as $p$ is slightly larger than $1/2$.  This is essentially a stronger statement of previous results obtained for infinite networks (see for example \cite{Acemoglu}), where the presence of signals with arbitrarily large informativeness, combined with the lack of individuals with unbounded influence, is identified as a sufficient condition for correct information aggregation. On the other hand, for $p < p_0$ the population reaches consensus on the wrong value of $X$ with non-zero probability.

\begin{figure}
\begin{center}
\resizebox{0.6\columnwidth}{!}{
  \includegraphics{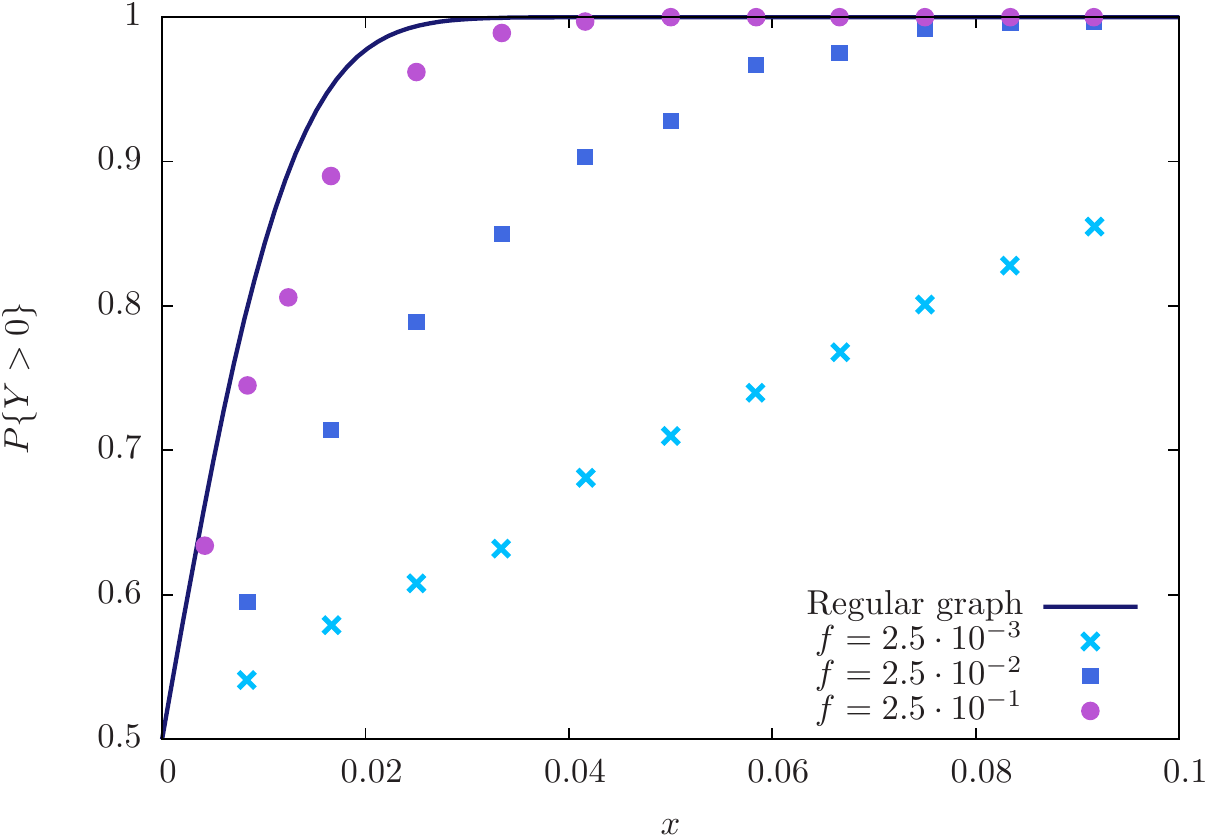}}
\end{center}
\caption{Probability of correct information aggregation as a function of the informativeness level of initial signals. The solid line refers to the case of a regular graph with $N=10^4$ (see equation \eqref{prob_clt}). The other data points refer to networks with $N=10^4$ and $c=4$ with different fractions of hubs $f$. As can be seen, ``perturbing'' of a regular graph with the introduction of a very small fraction of hubs seriously reduces the network's information aggregation performance. Increasing the fraction of hubs up to $f \lesssim c^{-1}$ progressively restores the regular graph levels of aggregation. Each data point was obtained by averaging over $10^4$ independent networks.}
\label{clt}
\end{figure}

A very interesting role in the information aggregation process is played by the fraction of hubs $f$. In Fig. \ref{clt}, one can see how, for a fixed system size $N$, the probability of correct information aggregation behaves when increasing the fraction $f$ of hubs in the network.
Also, it is rather interesting to compare such results with the information aggregation capabilities of a regular graph where all nodes have the same degree $c$. In such a case, one can immediately verify that the first eigenvector of the adjacency matrix is uniform with all components equal to $1/\sqrt{N}$, and the probability of the scalar product in equation \eqref{sc_prod_2} being positive simply reduces to the probability of the sum $\sum_{i=1}^N \theta_i(0)$ being positive. Therefore, one can compute the probability of correct information aggregation of a regular network with easy central limit theorem considerations, analogous to those already presented in this Section. Such a probability does not depend on $c$ and reads:
\begin{equation} \label{prob_clt}
P_\mathrm{RN} \{ Y > 0 \} = \frac{1}{2} \text{erfc} \left ( -x \sqrt{\frac{N}{2(1-x^2)}} \right ) \simeq \frac{1}{2} \text{erfc} \left ( -x \sqrt{\frac{N}{2}} \right ) \ ,
\end{equation}
where the last approximation holds for large values of $N$. As one can see, equation \eqref{prob_Y_3} reduces to the above expression for $f = c^{-1}$ (though numerically one does not find perfect agreement between the two, since equations \eqref{prob_Y_2} and \eqref{prob_Y_3} represent good approximations only for very low values of $f$). So, the lesson to be learned from the plots in Fig. \ref{clt} is twofold. First, one can see that as soon as a very small clique of uninformed hubs is introduced in a regular graph the overall population's ability to correctly aggregate information decreases sharply. This can be also understood by observing that the probability in equation \eqref{prob_Y_2} does not recover the regular network (RN) result \eqref{prob_clt} when considering vanishingly small fractions of hubs, i.e.:
\begin{equation} \label{diff_limits}
\lim_{f \rightarrow 0} P \{ Y > 0 \} \neq P_\mathrm{RN} \{ Y > 0 \} \ .
\end{equation}
On the other hand, whenever a clique of hubs is present in the network, then information aggregation can actually be improved by increasing the size of the clique itself, up to the point (for $f \simeq c^{-1}$) where the aggregation ability of the original regular graph can almost be reproduced.

Intuitively, the above findings can be altogether understood in the following terms. According to our setting, all hubs in the clique $\mathcal{H}$ are mutually connected and have a degree equal to $N_\mathcal{H} + c - 1$. This means that each hub has exactly $c$ neighbors outside $\mathcal{H}$, so that one can expect roughly $c N_\mathcal{H} = c f N$ nodes to fall within the clique's neighborhood $\partial \mathcal{H}$.  So, for very low values of $f$, $\partial \mathcal{H}$ contains a negligibly small number of nodes, which, however, will largely influence the initially uninformed hubs whenever they communicate for the first time. Given the small size of $\partial \mathcal{H}$, its initial state of knowledge will be much more sensitive to fluctuations in the initial signals distribution among agents. On the other hand, when $f \simeq c^{-1}$, the number of nodes in the neighborhood of $\mathcal{H}$ becomes of order $N$, hence much more robust with respect to fluctuations. 

In summary, the role of hubs in our model is subtle, as a handful of them is enough to heavily damage the good information aggregation properties of a population of equals (as modeled by a regular graph), whereas increasing their number also has ``healing'' effects which can restore such good properties.

\section{General dynamics}
\label{diff_dyn}

So far, we only have considered the most popular and widely used evolution rule for the information propagation on a network, i.e. the parallel dynamics introduced in equation \eqref{dynamics}. However, as already discussed in Section \ref{dyn_topology}, parallel dynamics represents one of the two extreme cases of the general dynamics \eqref{general_dyn}, according to which a fraction $\Phi$ of agents listens to their neighbors at each time step $t$, i.e. the case $\Phi = 1$. The other extreme case is the already mentioned RNS dynamics ($\Phi = 1/N$), according to which agents update their state of knowledge one at a time. 

\begin{figure}
\begin{center}
\resizebox{0.45\columnwidth}{!}{
  \includegraphics{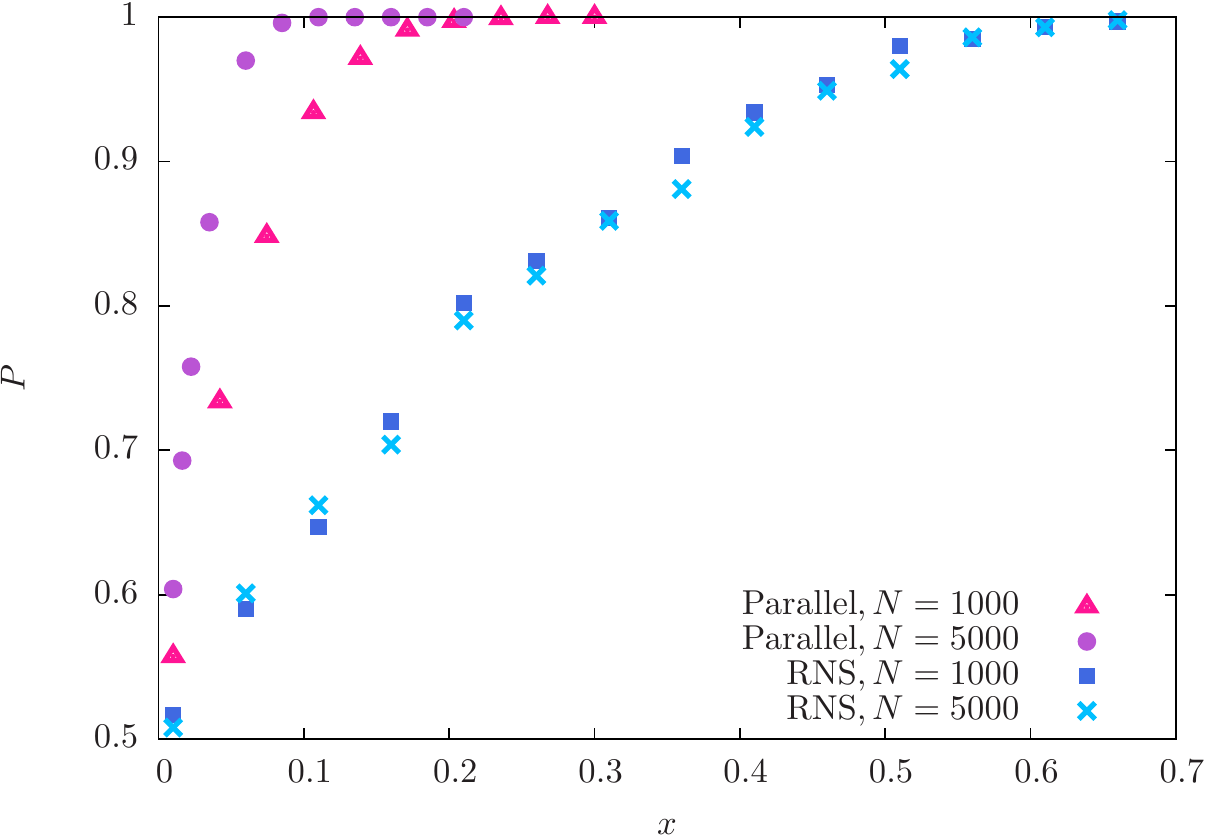}}
\hspace{1cm}
\resizebox{0.45\columnwidth}{!}{
  \includegraphics{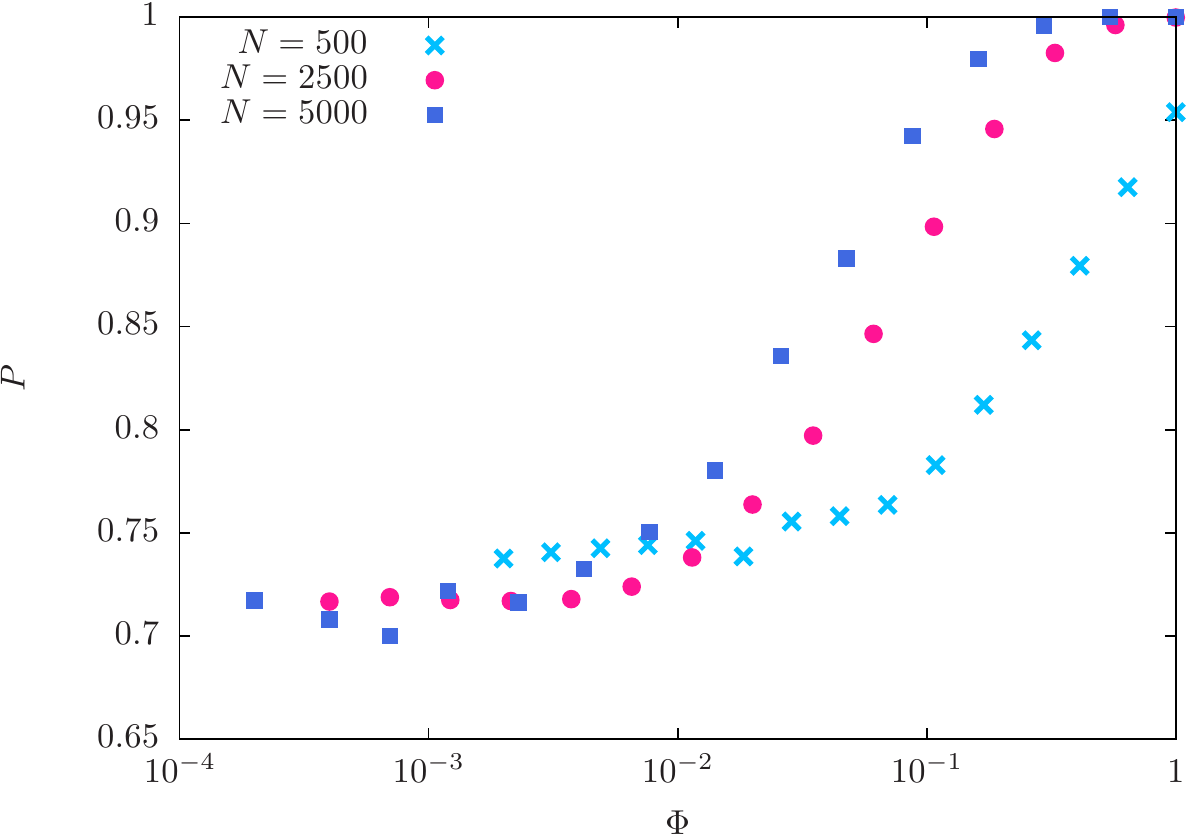}}
\end{center}
\caption{LEFT: Probability of correct information aggregation as a function of the informativeness level of initial signals. The different data sets refer to different types of dynamics run on networks with $N=10^3$ or $N = 5 \cdot 10^3$, $f=0.05$, and $c=4$. As can be seen, RNS dynamics does not show any significant dependence on the system size, and performs much worse than parallel dynamics at correctly aggregating information. RIGHT: Probability of correct information aggregation as a function of the fraction $\Phi$ of agents that listen to their neighbors at each time steps. The extreme cases $\Phi = 1/N$ and $\Phi = 1$ correspond, respectively, to RNS and parallel dynamics. All data were obtained for signal informativeness fixed as $x = 0.16$. In both plots all data points are obtained by averaging over $10^4$ independent network configurations.}
\label{transition}
\end{figure}

Numerical simulations highlight significant differences in a social network's ability to aggregate information correctly under parallel or RNS dynamics, the latter performing much worse than the former: as shown in the left panel of Fig. \ref{transition}, the probability of correct information aggregation under parallel dynamics outperforms the one obtained under RNS dynamics over a wide range of signal informativeness levels\footnote{We also performed numerical simulations under a random link sequential (RLS) dynamics, i.e. an information update rule according to which at each time step $t$ a link is randomly selected and the two nodes that share it exchange information. However, none of the simulations we performed highlighted any significant difference between such a dynamics and the parallel one.}. Moreover, results obtained via RNS dynamics show no relevant dependence on the system size $N$.
 
The above findings suggest to look for a transition in information aggregation as a function of the number of agents that update their state of knowledge at a given time step by letting the parameter $\Phi$ take values over the whole interval $[1/N,1]$. In the right panel of Fig. \ref{transition} we plot the probability $P$ of correct information aggregation as a function of $\Phi$ for different system sizes and a fixed informativeness level of the signals initially distributed to agents (the qualitative overall appearance of the results is not changed when considering different levels of informativeness).
As can be seen, for increasing values of $\Phi$ a transition is observed towards better information aggregation capabilities for all system sizes. This can essentially be interpreted in terms of the speed of information update. As one could expect, RNS dynamics is extremely slow compared to parallel dynamics (depending on the system size, we find on average that RNS dynamics reaches consensus in times that are 3-4 orders or magnitude larger than the ones required by parallel update), hence more prone to allow the spreading of misleading signals in the agents' initial distribution. On the other hand, parallel dynamics is fast, in such a way that in a few time steps each agent receives through his / her neighbors aggregated information coming from the whole network.

\section{Conclusions}
\label{conclusions}

In summary, we have presented a stylized dynamic network model of the information diffusion throughout a large society featuring a small fraction of uninformed leaders. The model's simplicity allows, in some cases, to make analytical considerations. Namely, when assuming all agents to simultaneously update their state of knowledge on a given issue, we are able to provide a closed-form expression for the probability of correct information aggregation as a function of the system size, i.e. the number of agents in the society, and the fraction of individuals playing the role of hubs. Our results partially overlap with previous works from the social learning literature in Economics, as we show that larger populations are better, on average, at aggregating information. On the other hand, we provide interesting novel results on the role played by the size of an uninformed \'elite, portrayed in our model by a clique of nodes that do not own any prior information on the issue being discussed by the population. First, we show a rather counterintuitive result, i.e. that increasing the relative size (compared to the overall population) of such uninformed \'elites actually helps the information aggregation process. Moreover, we show that letting the fraction of hubs go to zero does not recover the results obtained for the corresponding hub-free regular network.

Rather interestingly, we also show our model to be sensitive to the information update speed, as defined by the fraction of agents who simultaneously revise their information at each time step, by showing the existence of a transition towards better information aggregation capabilities when moving from the low speed towards the high speed regime.

\appendix

\section{Perturbative approximation of the top eigenvector $| \lambda_1 \rangle$}
\label{perturb}

When assuming hubs to be identified by nodes $1, \ldots, N_\mathcal{H}$, the network adjacency matrix $A$ takes the following block form:
\begin{equation} \label{adj}
A = \left ( \begin{array}{cc}
	I & G \\
	G^T & C \\
\end{array} \right ) \ .
\end{equation}
In the above equation, $I$ is an $N_\mathcal{H} \times N_\mathcal{H}$ block such that $I_{ij} = 1$ for $i \neq j$ and $I_{ii} = 0 \ \forall i$. The off-diagonal block $G$ is of size $N_\mathcal{H} \times (N - N_\mathcal{H})$, and it accounts for neighbors of the clique $\mathcal{H}$, i.e. $G_{ij} = 1$ for $i \in \mathcal{H}$ and $j \in \partial \mathcal{H}$, or vice versa, and zero otherwise. Lastly, the block $C$ is of size $(N - N_\mathcal{H}) \times (N - N_\mathcal{H})$, and it accounts for links between nodes that do not belong to $\mathcal{H}$.

Spectral properties of the adjacency matrix $A$, expressed in block form as in equation \eqref{adj}, can be deduced from standard perturbation theory. As a matter of fact, such a matrix can be decomposed as $A = A_\mathcal{H} + \tilde{A}$, where 
\begin{equation} \label{adj_decomp}
A_\mathcal{H} = \left ( \begin{array}{cc}
	I & 0 \\
	0 & 0 \\
\end{array} \right ) \ , \ \ \ 
\tilde{A} = \left ( \begin{array}{cc}
	0 & G \\
	G^T & C \\
\end{array} \right ) \ .
\end{equation}
For small values of $c$ (i.e. the degree of nodes outside of $\mathcal{H}$), the matrix $\tilde{A}$ above is sparse and can be interpreted as a perturbation to the matrix $A_\mathcal{H}$ describing the fully connected clique $\mathcal{H}$ plus a sea of $N-N_\mathcal{H}$ disconnected nodes.

Let us denote the eigenvalues and eigenvectors of the ``unperturbed'' adjacency matrix $A_\mathcal{H}$ as $\lambda_{i,\mathcal{H}}$ and $| \lambda_{i,\mathcal{H}} \rangle$. They fall within three categories:
\begin{itemize}
	\item The largest eigenvalue reads $\lambda_{1,\mathcal{H}} = N_\mathcal{H}-1$, and its normalized eigenvector $| \lambda_{i,\mathcal{H}} \rangle$ has the first $N_\mathcal{H}$ components equal to $1/\sqrt{N_\mathcal{H}}$ and the remaining $N-N_\mathcal{H}$ ones equal to zero.
	\item $\lambda_{i,\mathcal{H}} = -1$ for $i = 2, \ldots, N-N_\mathcal{H}$, with eigenvectors having non-zero components only in the first $N-N_\mathcal{H}$ sites.
	\item $\lambda_{i,\mathcal{H}} = 0$ for $i = N-N_\mathcal{H}+1, \ldots, N$, with eigenvectors that can simply be chosen as having all components equal to zero except for the $i$-th component being equal to one.
\end{itemize}
Let us then approximate the first eigenvector of the full adjacency matrix $A$ as
\begin{equation} \label{approx_eig}
| \lambda_1 \rangle \simeq | \lambda_{1,\mathcal{H}} \rangle + | \lambda_1^\prime \rangle + | \lambda_1^{\prime \prime} \rangle \ ,
\end{equation}
where $| \lambda_1^\prime \rangle$ and $ | \lambda_1^{\prime \prime} \rangle$ denote the first and second order corrections, respectively, to the unperturbed eigenvector $| \lambda_{1,\mathcal{H}} \rangle$. The first order correction only involves neighbors of the clique $\mathcal{H}$, and it reads
\begin{equation} \label{first_ord_vect}
| \lambda_1^\prime \rangle = \sum_{j > 1} \frac{\langle \lambda_{j,\mathcal{H}} | \tilde{A} | \lambda_{1,\mathcal{H}} \rangle}{\lambda_{1,\mathcal{H}} - \lambda_{j,\mathcal{H}}} | \lambda_{j,\mathcal{H}} \rangle =
\frac{1}{\sqrt{N_\mathcal{H}}(N_\mathcal{H} - 1)} \sum_{j \in \partial \mathcal{H}} n_j^\prime | \lambda_{j,\mathcal{H}} \rangle \ ,
\end{equation}
where $n_i^\prime = \sum_{j \in \mathcal{H}} a_{ij}$ represents the number of neighbors that node $i$ has within the clique $\mathcal{H}$. The second order correction\footnote{Second order corrections also involve the first the first $N_\mathcal{H}$ components of $\lambda_{1,\mathcal{H}}$. However, such corrections are irrelevant to our analysis, so we can safely neglect them.} involves neighbors of the nearest neighbors of the clique $\mathcal{H}$:
\begin{eqnarray} \label{second_ord_vect}
| \lambda_1^{\prime \prime} \rangle &=& \sum_{j > 1} | \lambda_{j,\mathcal{H}} \rangle \sum_{\ell > 1} \frac{\langle \lambda_{j,\mathcal{H}} | \tilde{A} | \lambda_{\ell,\mathcal{H}} \rangle \langle \lambda_{\ell,\mathcal{H}} | \tilde{A} | \lambda_{1,\mathcal{H}} \rangle}{(\lambda_{1,\mathcal{H}} - \lambda_{j,\mathcal{H}}) (\lambda_{1,\mathcal{H}} - \lambda_{\ell,\mathcal{H}})} \\ \nonumber
&=& \frac{1}{\sqrt{N_\mathcal{H}} (N_\mathcal{H} - 1)^2} \sum_{j \in \partial(\partial \mathcal{H})} n_j^{\prime \prime} | \lambda_{j,\mathcal{H}} \rangle \ ,
\end{eqnarray}
where $\partial (\partial \mathcal{H})$ denotes the set of next to nearest neighbors of the clique $\mathcal{H}$, whereas $n_i^{\prime \prime} = \sum_{j \in \partial \mathcal{H}} a_{ij}$ is the number of neighbors that node $i$ has amongst neighbors of the clique $\mathcal{H}$. In order to perform exact calculations up to second order, one should in principle compute the expected number of nodes belonging to $\partial \mathcal{H}$ and $\partial (\partial \mathcal{H})$, and the expected values of the quantities $n_j^\prime$ in \eqref{first_ord_vect} and $n_j^{\prime \prime}$ in \eqref{second_ord_vect} by averaging over all possible network configurations built as explained in Section \ref{dyn_topology} for given $N$, $N_\mathcal{H}$ and $c$. However, in order to keep things simple, let us just assume that each node in $\partial \mathcal{H}$ has just one neighbor in the clique $\mathcal{H}$, and, in a similar fashion, that each node in $\partial (\partial \mathcal{H})$ has just one neighbor in $\partial \mathcal{H}$, which amounts to posing $n_i^\prime = 1$, $\forall i \in \partial \mathcal{H}$, and $n_j^{\prime \prime} = 1$, $\forall j \in \partial (\partial \mathcal{H})$. Clearly, both such approximations work well as long as the number of nodes in $\mathcal{H}$ is small compared to $N$, i.e. for $f \ll 1$ where $f = N_\mathcal{H} / N$. 

According to the above approximations, the $N - N_\mathcal{H}$ nodes not belonging to $\mathcal{H}$ yield the following components in $| \lambda_1 \rangle$, as computed with equation \eqref{approx_eig}:
\begin{itemize}
	\item $c N_\mathcal{H}$ components (i.e. the number of nodes in $\partial \mathcal{H}$) equal to $1/(\sqrt{N_\mathcal{H}} (N_\mathcal{H} - 1)) \simeq 1 / N_\mathcal{H}^{3/2}$
	\item $c(c-1) N_\mathcal{H}$ components (i.e. the number of nodes in $\partial (\partial \mathcal{H})$) equal to $1/(\sqrt{N_\mathcal{H}} (N_\mathcal{H} - 1)^2) \simeq 1 / N_\mathcal{H}^{5/2}$
	\item $N - (1+c^2) N_\mathcal{H}$ components equal to zero.
\end{itemize}
Therefore, the mean $\mu_\lambda$ and standard deviation $\sigma_\lambda$ (see equation \eqref{stat_Y}) can be computed as follows:
\begin{eqnarray} \label{stat_lambda_2}
\mu_\lambda &=& \frac{1}{N - N_\mathcal{H}} \left ( \frac{c}{\sqrt{N_\mathcal{H}}} + \frac{c(c-1)}{N_\mathcal{H}^{3/2}} \right ) \\ \nonumber
\sigma_\lambda^2 &=& \frac{1}{N - N_\mathcal{H}} \left ( \left ( \frac{1}{N_\mathcal{H}^{3/2}} - \mu_\lambda \right )^2 c N_\mathcal{H} + \left ( \frac{1}{N_\mathcal{H}^{5/2}} - \mu_\lambda \right )^2 c(c-1) N_\mathcal{H} +
\mu_\lambda^2 \left ( N - (1+c^2) N_\mathcal{H} \right )  \right ) \ ,
\end{eqnarray}
and the approximations in equation \eqref{stat_lambda} can be immediately derived as leading order results in $N$ of the above expressions.

\end{document}